# OCTAVA: an open-source toolbox for quantitative analysis of optical coherence tomography angiography images

**Gavrielle R. Untracht,[1,2,*] Rolando Matos,[3] Nikolaos Dikaios,[4] Mariam Bapir,[3] Abdullah K. Durrani,[2] Teemapron Butsabong,[3] Paola Campagnolo,[3] David D. Sampson,[2] Christian Heiss[3,5] and Danuta M. Sampson[3,6]**

[1]*Optical+Biomedical Engineering Laboratory, School of Electrical, Electronic and Computer Engineering, The University of Western Australia, 35 Stirling Highway, Perth WA 6009, Australia*

[2]*Surrey Biophotonics, Advanced Technology Institute, School of Physics and School of Biosciences and Medicine, University of Surrey, Guildford, Surrey GU2 7XH, United Kingdom*

[3]*Department of Biochemical Sciences and Department of Clinical and Experimental Medicine, School of Biosciences and Medicine, University of Surrey, Guildford, Surrey GU2 7XH, United Kingdom*

[4]*Mathematics Research Center, Academy of Athens, Athens 106 79, Greece*

[5]*Surrey and Sussex Healthcare NHS Trust, East Surrey Hospital, Redhill, Surrey, RH1 5RH, United Kingdom*

[6]*Surrey Biophotonics, Centre for Vision, Speech and Signal Processing and School of Biosciences and Medicine, The University of Surrey, Guildford, GU2 7XH United Kingdom*

Corresponding author:

Email: gavuntracht@gmail.com

**Abstract**

Optical coherence tomography angiography (OCTA) performs non-invasive visualization and characterization of microvasculature in research and clinical applications mainly in ophthalmology and dermatology. A wide variety of instruments, imaging protocols, processing methods and metrics have been used to describe the microvasculature, such that comparing different study outcomes is currently not feasible. With the goal of contributing to standardization of OCTA data analysis, we report a user-friendly, open-source toolbox, OCTAVA (OCTA Vascular Analyzer), to automate the pre-processing, segmentation, and quantitative analysis of *en face* OCTA maximum intensity projection images in a standardized workflow. We present each analysis step, including optimization of filtering and choice of segmentation algorithm, and definition of metrics. We perform quantitative analysis of OCTA images from different commercial and non-commercial instruments and samples and show OCTAVA can accurately and reproducibly determine metrics for characterization of microvasculature. Wide adoption could enable studies and aggregation of data on a scale sufficient to develop reliable microvascular biomarkers for early detection, and to guide treatment, of microvascular disease.

## 1. Introduction

The microcirculation comprises the smallest elements of the circulatory system, a dense network of arterioles, capillaries, venules, and lymphatic vessels with a diameter of less than 150 μm [1]. These small vessels, the microvasculature, account for about 99% of blood vessels in adults and play a key role in oxygen transport and nutrient delivery to the tissue, as well as in waste and carbon dioxide removal. Many studies have suggested that dysfunction in the microcirculation may be as important as dysfunction in larger blood vessels in the context of cardiovascular pathophysiology [2]. However, due to the lack of accessible/available diagnostic modalities tailored to the microvasculature, knowledge of its precise role in disease pathogenesis and progression is lacking and no specifically microvascular therapies are currently available [3]. With the introduction of optical coherence tomography angiography (OCTA), the study and visualization of the microvasculature is well underway in the human retina and, to a lesser extent, in human skin and animal models of human disease [4–6]. Standard "structural" OCT uses interference to measure the echo arrival time and intensity of backscattered light to generate cross-sectional and volumetric images of optical scattering in tissue [7]. By using differences between sequential OCT images (usually cross-sectional B scans) at the same location, caused by motion in the sample, OCT volumetric images can be processed to generate volumetric or two-dimensional representations of blood flow, angiograms, which distinguish structures containing flowing blood from the surrounding static tissue [8]. OCTA angiograms typically comprise *en face* maximum intensity projection (MIP) images, where MIPs have been shown to be superior to other projections [9]. OCTA is an ideal technique for clinical imaging because it does not require exogenous contrast agents; contrast is generated from blood flow, largely from red blood cells. Based on OCTA images, it is possible to describe the microvascular network architecture and vessel morphology and introduce, describe, and quantify various metrics to characterize them. Some commonly used metrics include: vessel area density, total vessel length, and mean vessel diameter, all within a region of interest [10]. It is also possible to undertake a temporal analysis of pulsatility towards characterizing pathophysiology or arterial stiffening [11], or measure microvascular response to external stimuli, such as heating or pressure, to enhance diagnostic accuracy [12,13].

There are a number of commercial and non-commercial, clinical and laboratory-based OCTA instruments and systems that enable visualization and characterization of microvasculature. These systems invariably use different imaging engines, imaging protocols, image processing methods and metrics to describe the microvasculature. As a result, comparing study outcomes obtained with different instruments and generating large databases across instruments and institutions is not possible. Indeed, it has been proven that quantifiable metrics obtained using different OCTA instruments are not readily comparable [14,15]. As a more specific example, differences in binarization thresholding methodologies have been proven to significantly influence the quantification of OCTA metrics in healthy eyes [16,17]. Standardized approaches to process and analyze OCTA images are needed. It is crucial that metrics generated by the research and clinical community are accurate, consistent and reliable if these metrics are to be used to generate agreed-upon biomarkers associated with diagnosis, monitoring, and treatment guidance of disease [18,19].

Fortunately, there is a growing interest within the research community in delivering universal tools that enable visualization and quantification of microvasculature. A number of tools originally developed for microscopy can be applied to OCTA images [20–24], even if none have yet been optimized for OCTA. Some toolboxes limited in scope, in terms of application or instrument, have been made available for evaluation of OCTA images [25–27], with differences in processing methods and quantitative metrics. The lack of consistency between such toolboxes limits their ability to compare results acquired with different instruments, in different studies, or in different institutions, and, thus, the ability to draw conclusions about the biological relevance of the results, pointing to the need for a more versatile and universal alternative.

We have developed and validated a new, easy to use, open-source software toolbox, OCTA Vascular Analyzer (OCTAVA), towards standardization of OCTA image analysis and characterization. Standardization implies choices of the free parameters in such a toolbox. We have identified from the extensive existing literature the best algorithms for OCTA image analysis and characterization. We compare in detail the performance of the five best performing segmentation algorithms and select eight metrics by assessing quantitatively and qualitatively their capability to visualize and characterize microvasculature. Finally, we implemented the optimal workflow in OCTAVA: i) pre-processing and segmentation of *en face* OCTA images; ii) identification of the vascular network and nodes using skeletonization; iii) automated measurements of the length, diameter, and tortuosity of the each identified vessel; and iv) generation of the microvascular network architecture and vessel morphology metrics: vessel area density, vessel length density, mean and median vessel diameter, total vessel length, branchpoint density, mean tortuosity and fractal dimension. Using images from various sources, we demonstrate that OCTAVA can accurately and reproducibly determine metrics for characterization of microvasculature independent of instrument and application.

## 2. Methods

### 2.1 OCTA imaging and human subjects

OCTA images used for toolbox development, validation and optimization were acquired using the multi-beam VivoSight Dx (*Michelson Diagnostics Ltd*, Maidstone, Kent, UK). This swept-source OCT instrument allows imaging of microvasculature of the skin with 20 kHz line-scan imaging speed and provides an imaging resolution of 5.5 μm and 7.5 μm in the axial and transverse directions, respectively. Eight healthy individuals aged between 25 and 62 years old were enrolled in this study, for which a favorable ethical opinion was issued by the University of Surrey Ethics Committee (FHMS 19-20 060). Written, informed consent was obtained from all participants in adherence to the Declaration of Helsinki. For all participants, the skin on the dorsum of the right hand, between the thumb and forefinger, was imaged. The handheld OCT probe was positioned for imaging on the skin through a plastic cap to reduce motion artifacts and maintain a constant distance between the imaging probe and the skin. Volumetric images were acquired over a 5 mm by 5 mm area with spatial sampling of 4.4 μm along the fast axis and 41 μm along the slow axis. The built-in software "VivoTools" was used to generate volumetric representations of blood flow based on pairs of cross-sectional B-scans. A surface detection algorithm was then applied in MATLAB 2020a (*The MathWorks, Inc.,* Natick, Massachusetts, USA) to identify and flatten the curved skin surface. The top 175-μm-deep section of the image was removed to exclude the avascular epidermis and dermal papillae. Dermal papillae were excluded since vessels in the dermal papillae are oriented along the imaging axis; thus, they appear as dots and their interconnectivity with the network is difficult to analyze. The depth range of our images corresponds with the superficial plexus, a dense network of vessels oriented parallel to the skin surface. Two-dimensional (2D) *en face* OCTA angiograms were generated from the volumetric angiograms in MATLAB using a maximum intensity projection (MIP) over a physical thickness of 500 µm in depth (unless otherwise specified), assuming an average group refractive index of 1.4.

## 2.2 Comparison and choice of segmentation algorithms

Previous work has noted significant variability in resulting metrics using different segmentation algorithms [28–31]. As such, we have evaluated the performance of five different image segmentation algorithms to determine the optimal approach for OCTA MIP images in skin: a global thresholding approach using the built-in "convert to mask" function in ImageJ; k-means [32]; iterative self-organizing data analysis technique (ISODATA) [33]; adaptive thresholding [34]; and fuzzy thresholding [35]. These algorithms were selected by reviewing viable approaches in the retinal OCTA literature based on the number of papers that use them and the reported accuracy of the results. Out of tens of algorithms, we selected five from a preliminary assessment as demonstrating the best performance [28–30].

The five segmentation algorithms were first optimized empirically; any free parameters were adjusted to maximize recognition of vessel structures and minimize misidentification of background signals as vessels, assessed manually. For the ImageJ approach, the global threshold is calculated automatically based on the mean pixel intensity in the image; pixels with a value greater than the mean are vessels. The fuzzy thresholding algorithm is a spatial aggregation method which automatically fits a set of probability density functions to the histogram of the image to determine the classification of each pixel. K-means is an unsupervised algorithm that clusters the image into k-clusters (where each voxel belongs to the cluster with the nearest mean). Adaptive thresholding and ISODATA are both local adaptive thresholding algorithms that use a specified window size to generate a local threshold for each pixel. These algorithms have difficulty in identifying larger uniform areas, so a large window size was required to minimize misidentification of uniform areas.

The performance of the five segmentation algorithms was compared using selected OCTA MIP images with different densities of vascular structures to assess how accurately the binarized image represented the OCTA data. Even if it is currently used as a gold standard in many other works, we opted not to use manual segmentation of the vascular network as a ground truth because it generally has poor accuracy and precision compared with automatic segmentation [29], and the results vary based on the experience of the person performing the segmentation [36]. Instead, the algorithms were compared empirically and quantitatively using the metrics of vessel area density (VAD), network connectivity factor (CF), and repeatability. VAD is described in detail in **Section 2.5**. The network connectivity factor, CF, quantifies the percentage of identified vessels that are connected to the main network using the formula $CF = 1 - \frac{S_i}{S_t}$, where $S_i$ is the number of isolated elements in the image (segments with neither endpoint connected to the network) and $S_t$ is the total number of identified vessel segments in the image. Repeatability was assessed using the standard deviation of VAD measured from OCTA MIP images acquired sequentially at the same location without adjusting the placement of the handheld probe.

## 2.3 OCTAVA development environment

Software development was mainly undertaken in MATLAB 2020b (*The MathWorks, Inc.,* Natick, Massachusetts, USA) for ease of optimization, graphical user interface (GUI) development, and customization. We utilize the ImageJ-MATLAB package to access ImageJ (*US National Institutes of Health*, Bethesda, Maryland, USA) libraries from MATLAB, which allows us to fully integrate the image processing capabilities in ImageJ and the automation and data management tools in MATLAB [37]. This approach allows us to build on existing functionality within ImageJ while giving us the flexibility to investigate different segmentation algorithms and collation of metrics, which is more straightforward to implement in MATLAB. Our OCTAVA software adopts and integrates the Angiogenesis Analyzer [22] developed within ImageJ [38] with several important modifications. First, we have developed and optimized image pre-processing and segmentation in MATLAB, which ensures good quality recognition of vessels in OCTA MIP images. Second, we have expanded the available quantitative metrics describing vasculature beyond those generated by Angiogenesis Analyzer to include additional information about the length, diameter, and tortuosity of individually identified vessels and the interconnectivity of the network, which enables us to perform more sophisticated statistical analyses on vessel data and to generate additional metrics beyond those available with other software. Finally, we have developed a dedicated GUI, shown in **Fig 1,** that allows automated segmentation, identification of vessel components, and compilation of results from within a single interface. Because our software is open source and developed in MATLAB, it can easily be modified to adapt to the needs of the research and clinical community while still being easy to use without modification. A MATLAB license is required to modify the software; however, a standalone version of OCTAVA is available that can be used without MATLAB.

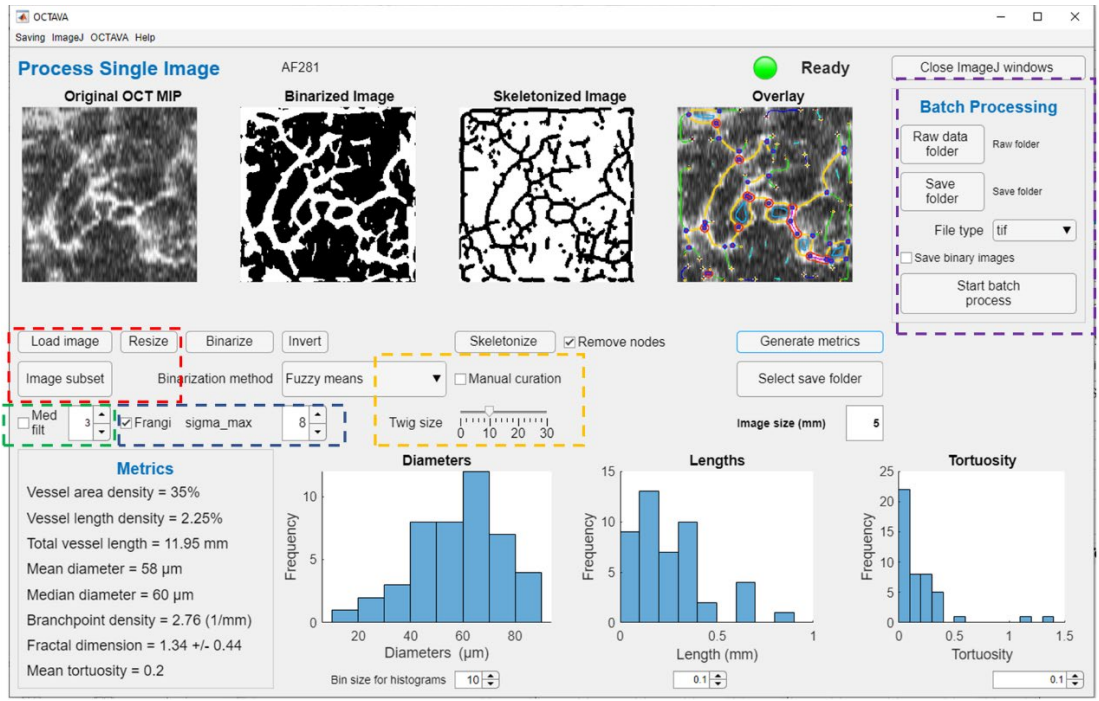

**Fig 1. Labelled image of OCTAVA graphical user interface.** Colored boxes indicate important user controls for optimization of image processing. Red box: the user can modify the image by down sampling or selecting a subregion of the full image for faster processing. Green box: median filter controls. Blue box: Frangi filter controls. Yellow box: vascular analysis controls. Purple box: batch processing controls.

## *2.4 OCTAVA software*

### OCTAVA GUI

The OCTAVA GUI provides a single interface for users to process OCTA MIP images and to view results. The user has the option of selecting a region of interest (a subregion of the image) for processing (rectangular or circular) and rescaling the number of pixels in the image – either to increase processing speed or to improve the precision of the metrics (red box in **Fig 1**). OCTAVA has two different operating modes: one which allows the user to process individual images with the aim of optimizing the analysis protocol for specific image types, and a batch-processing mode which allows the user to analyze multiple images without additional user input (purple box in **Fig 1**). In both cases, the same image processing workflow is utilized; each of these steps is described in detail in the following sections and visualized in **Fig 2**. Similarly to REAVER [23], we give the user an option for manual curation to allow for correction in the case of improper automated segmentation or vessel identification (yellow box in **Fig 1**). Manual curation is performed by manually adding or removing items from the list of generated regions of interest (ROIs) within ImageJ. Many metrics have been implemented to enable the identification of those most meaningful for use in diagnosis and monitoring of disease and response to treatment. These will be discussed further in **Section 2.5**.

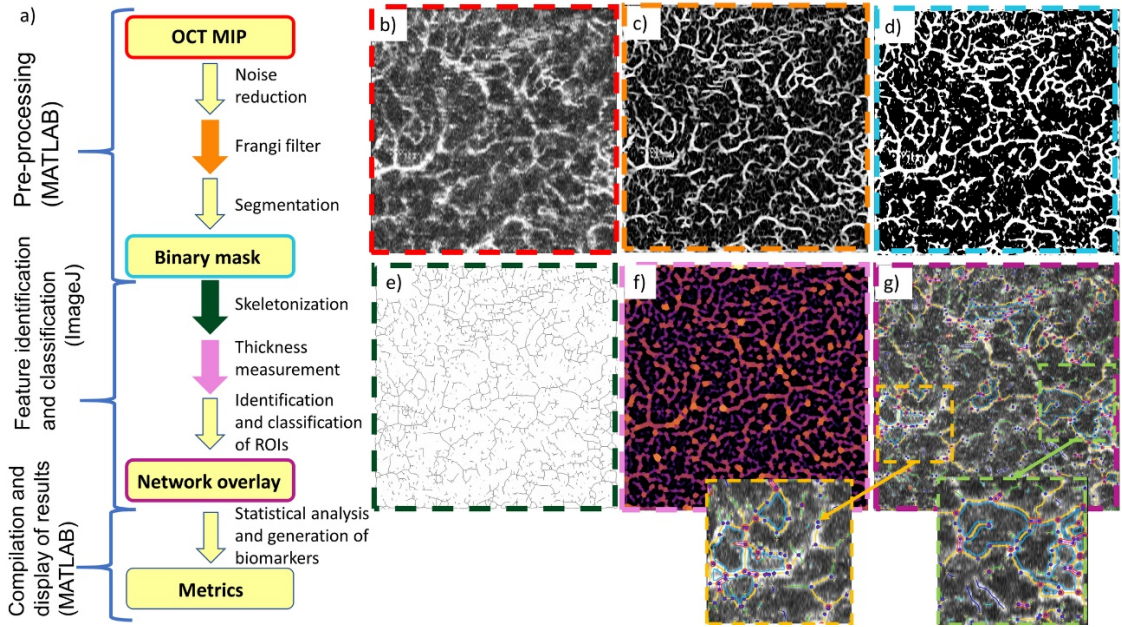

**Fig 2. Flowchart of software workflow.** (a) and example images representing each step in the workflow (b-g). Colored borders in (b-g) correspond to the step with the matching-colored arrow or box in (a). OCTA MIP image (b) is loaded into OCTAVA. Images are first pre-processed in MATLAB to remove noise and enhance the signal intensity of vessel-like structures using a Frangi filter (c). Images are then segmented using the chosen segmentation algorithm to generate a binary mask (d). The image is then sent to ImageJ, where the network is skeletonized (e). The thickness and interconnectivity are measured (f), and the ROIs and network elements are identified and classified based on their thickness and interconnectivity. An overlay image is generated (g) which helps the user confirm that an accurate map of the network architecture was measured. Colors in (f) correspond to the local vessel thickness. Colors in (g) correspond to different architectural components of the network including segments (yellow lines), branches (green lines), mesh regions (light blue lines), isolated elements (dark blue lines), and nodes (red and blue circles). The insets in (g) allow a closer examination of the identified network elements. Finally, the outputs of the ImageJ analysis are sent to MATLAB, which generates and compiles metrics of the network. The large images in (b-g) are 5 mm x 5 mm. The insets in (g) are 1.6 mm x 1.6 mm.

### Image processing workflow

OCTA MIPs are pre-processed using MATLAB to generate a binary mask using the following procedure. OCTA MIPS are first assessed by the user as having sufficient image quality to proceed. A 2D *Frangi "vesselness" filter* is applied since the objects of interest are blood vessels [39] (**Fig 2c**). The Frangi filter is commonly used in analysis of angiography images since it reduces the impact of intensity variations along a vessel and suppresses background noise, thereby improving image segmentation [40]. The 2D "vesselness" filter is used rather than a 3D filter for accessibility, since many OCTA instruments limit access to the volumetric data (volumetric data is acquired, but often not made readily available to the user). A multi-scale approach is used to minimize artificial vessel dilation (optimization of the filter is discussed further in **Section 3.2**). The Frangi filter enhances the signal-to-noise ratio (SNR) of the image by reducing noise and intensity variations along the length of vessels. The OCTA MIPs are segmented into "vessel" and "not vessel" regions (**Fig 2d**) using the optimal segmentation algorithm as determined by our analysis in Section **3.2**. The binarized image is then skeletonized using a 3D thinning algorithm in ImageJ (**Fig 2e**) [40,41] and a heatmap of vessel thickness is generated (**Fig 2f**) [42]. The algorithm for measuring local thickness is described in **Section 2.5**. Nodes are identified based on their diameter relative to other nearby structures and removed so they do not impact diameter data. The dimensions of individual vessel segments are recorded and written to a spreadsheet for post-processing and compilation of results using MATLAB.

### Identification of network elements

OCTAVA identifies different elements that make up the network and displays this information in a color-coded overlay of the skeletonized image on top of the original image (**Fig 2g**). Linear regions are separated into three categories based on their interconnectivity with the network: segments, colored yellow, are connected to other vessels at both endpoints (via nodes); branches (green) are connected to another vessel on one end but have one free endpoint; isolated elements (dark blue) are not connected to other vessels on either end. Isolated elements below a certain diameter (specified using the "twig size" control in the OCTAVA GUI) are likely noise and are excluded from the analysis. Nodes, marked in the image with blue and red circles, are counted but are removed from the diameter measurements to avoid artificially overestimating vessel diameters. Mesh regions, which are areas completely enclosed by the vessel network, are identified in cyan. Examples of each of these structures can be seen in the insets in **Fig 2g**.

## 2.5 Quantitative metrics for characterization of microvascular network

The microvascular network forms a complex architecture of interconnected vessels. The complex spatial network undergoes remodeling in embryonic and adult tissues, as well as in quiescent and pathological states. Such complexity signifies that full characterization of changes in vessel morphology will require multiple metrics. We chose eight metrics to comprehensively characterize microvascular network architecture based on an extensive qualitative review of the literature, balancing coverage and overlap of characteristics [13,43–47]. These metrics allow users to obtain comprehensive information about the microvasculature [48]. **Table 1** and **Fig 3** give an overview of our selected metrics, how they are calculated, and indications of biological relevance. We note that the potential biological relevance is not exhaustive – this is an open area of research which will be facilitated by our software – and is included here as a brief justification of the breadth of metrics we chose to implement. All metrics are based on the 2D OCTA MIP image.

**Table 1. Microvascular network architecture quantitative metrics.**

| Metric | Unit[a] | Description | Potential biological relevance |
|---|---|---|---|
| Vessel area density (VAD) | % | Perfused blood vessel area in binarized OCTA MIP image divided by total image area | Indicates utilization of microvessels. Increased VAD is associated with angiogenesis |
| Vessel length density (VLD) | % | Total length of perfused vessels measured along the vessel centerline divided by the total image area. Length measurements are calculated from the skeletonized OCTA MIP image | Indicates oxygenation/nutrient delivery dysfunction. Increased VLD is associated with angiogenesis |
| Average and distribution of vessel diameter | μm | Diameter measurements acquired using a local thickness algorithm in ImageJ applied to the binarized OCTA MIP | Reveals information about dilation and regression; the distribution allows further differentiation between changes in diameter and number of perfused vessels |
| Average and distribution of vessel length | mm | Length of each identified vessel segment along the centerline of the vessel calculated from the skeletonized OCTA MIP image | Network interconnectivity and branching patterns indicates oxygenation/nutrient delivery dysfunction |
| Average and distribution of tortuosity | 1 | Segment length along the centerline divided by chord length for each identified vessel calculated from the skeletonized OCTA MIP image | Increased blood vessel tortuosity can indicate pathological microvascular remodeling and/or ischemia |
| Branchpoint density | Nodes/mm | Number of identified nodes divided by total vessel length | Network interconnectivity-based indicator of resistance to blockage or occlusion and flow dynamics |
| Fractal dimension | 1 | Indication of how the network fills space on variable length scales, calculated using the box counting method | Indicator of branching architecture within the network; indicator of tortuosity and microvascular remodeling |

[a]A unit of 1 indicates a dimensionless metric.

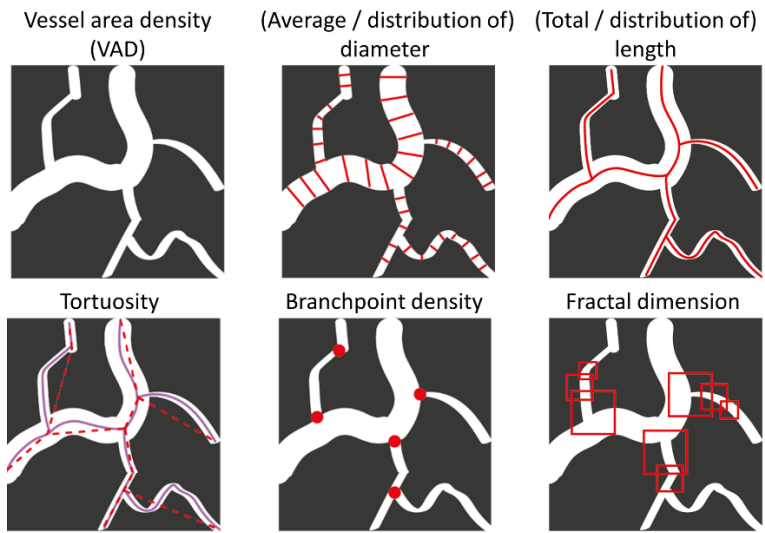

**Fig 3. Visual representation of metrics.**

VAD is calculated from the binary mask by counting the total area of the identified vessels (white pixels) divided by the total number of pixels. The vessel length density (VLD) is calculated by counting the total number of black pixels in the skeletonized image, which represents the sum of the length of all the vessels (total vessel length) and dividing by the total number of pixels in the image. For the purposes of the VLD calculation, a vessel diameter of 1 pixel is assumed in order to have units of %. For other purposes, vessel diameters are calculated using the local thickness algorithm in ImageJ applied to the binarized OCTA MIP image [42]. This algorithm fits circles into the vessels; the diameter of each circle corresponds to individual measurement of vessel diameter and an average diameter is calculated for each identified vessel segment. Overall mean and median diameter values are calculated as the average diameter of the vessel segments. The fractal dimension is calculated using the box counting method [49,50], which consists of dividing the image into square boxes of equal size and counting the number of boxes which contain part of the vessel network. This process is iteratively repeated with boxes of different sizes and the number of counted boxes is plotted against the box size on a log-log plot. The average negative slope of the resulting line is the fractal dimension, defined by $s = -\frac{\log_{10} N(l)}{\log_{10} l}$, where $l$ is the box length and $N(l)$ is the number of boxes which contain part of the vessel network. The resulting value will be between 0 and 2. True fractals will be linear throughout the whole plot, with values approaching 2 indicating more fractal-like branching structures. Quasi-fractal images may be linear over a small region; therefore, fractal dimension is typically reported with the standard deviation which gives an indication of the linearity; a lower standard deviation corresponds with more fractal-like structures. Branchpoint density is calculated by counting the number of nodes in the image and dividing by total vessel length. Tortuosity is calculated from the skeletonized image using the arc length-over-chord ratio, where the arc length is the total length of the identified segment between two nodes, $L_s$, and the chord length is the length of a straight line between the two endpoints, $L_C$, so that $Tortuosity = \frac{L_S}{L_C} - 1$ [51]. Additional methods for quantifying tortuosity for different applications will be incorporated into future versions of OCTAVA.

In addition to the average length, diameter, and tortuosity measurements, we present histograms of the distribution of these values within the image. Previous work has shown that the distribution of these values can provide further information, for example, for distinguishing changes in total number of perfused vessels versus vessel dilation for increased VAD [52].

*2.6 Evaluation of OCTAVA*

The OCTA images of human skin described above were used to evaluate various pre-processing filters and segmentation algorithms to optimize the OCTAVA workflow. Additionally, we have used OCT images of intralipid flow in microfluidic devices obtained with a spectral domain OCT system described elsewhere [53] and a simulated OCTA MIP image to evaluate the accuracy of algorithms for metrics calculation. Manual characterization of the microfluidic device was performed in ImageJ. Exact metric values for the simulated OCTA MIP image were generated using *a priori* knowledge of the image structure. As well, to demonstrate the utility of the software in various types of OCTA images, we have used OCTA (MIP) images from several previously published works. These include: OCTA images of human skin collected with a laboratory-based polarization-sensitive OCT system [54], OCTA images of mouse ear skin acquired with the Telesto spectral domain OCT system (*Thorlabs*, Newton, NJ, USA) [55], and images of the human retina obtained using the RTVue XR Avanti (*Optovue, Inc.,* Fremont, CA, USA) [56].

## 3. Results

### 3.1 Optimization of Frangi "vesselness" filter

We investigated the parameters of the *Frangi "vesselness" filter* to optimize the visibility and characterization of vessels within the image [39,57]. The Frangi filter selectively enhances the intensity of individual pixels based on their "vesselness," which is quantified using the eigenvalues of a local Hessian matrix $H(x, y, \sigma) = \sigma^2 I(x, y) \frac{\partial^2}{\partial x \partial y} G(x, y, \sigma)$, where $I(x, y)$ is the intensity of a pixel in the image and $G(x, y, \sigma)$ is a 2D Gaussian kernel: $G(x, y, \sigma) = \frac{1}{2\pi\sigma^2} e^{\frac{-(x^2+y^2)}{2\sigma^2}}$. In this formula, $\sigma$ represents the expected vessel diameter. Optimization of the Frangi filter for each image type based on the expected diameter of vessels is an important step since it can impact the measured diameter. Similarly to other works [40,58], we have used a multi-scale approach which allows us to enhance a range of vessel diameters by iterating over a range of $\sigma$ values from 0 to $\sigma_{max}$, where $\sigma_{max}$ is the expected maximum vessel diameter. Selection of too large a value for $\sigma_{max}$ will lead to artificial vessel dilation whereas underestimation will lead to insufficient SNR enhancement. While the range of $\sigma$ values can be estimated based on *a priori* knowledge of the expected range of vessel diameters within an image and verified empirically, we determined that measuring the apparent full width at half maximum (FWHM) of the vessel diameter directly from the OCT image provided a more quantitative benchmark for optimization. For example, [58] used values in the range 1-10 for $\sigma$ corresponding to vessel diameters in the range 10-100 μm in skin in the crook of the elbow and knee. Through our analysis, we determined that $\sigma$ values in the range 1-8, corresponding to vessel diameters in the range 10-80 μm, is more appropriate for our measurement location on the hand. The results of our analysis are shown in **Fig 4.** The thick black line indicates the intensity profile measured directly from the OCTA MIP image. For all values of $\sigma_{max}$, the background signal level outside of the vessel has been reduced, but for $\sigma_{max} \leq 4$, the signal intensity of the vessel itself is also reduced. For larger values of $\sigma_{max} \geq 12$, the apparent vessel diameter is overestimated. By inspection, we chose the median value of $\sigma_{max} = 8$ as representing the best trade-off between accurate vessel diameter and vessel enhancement for the range of vessel diameters present in our images.

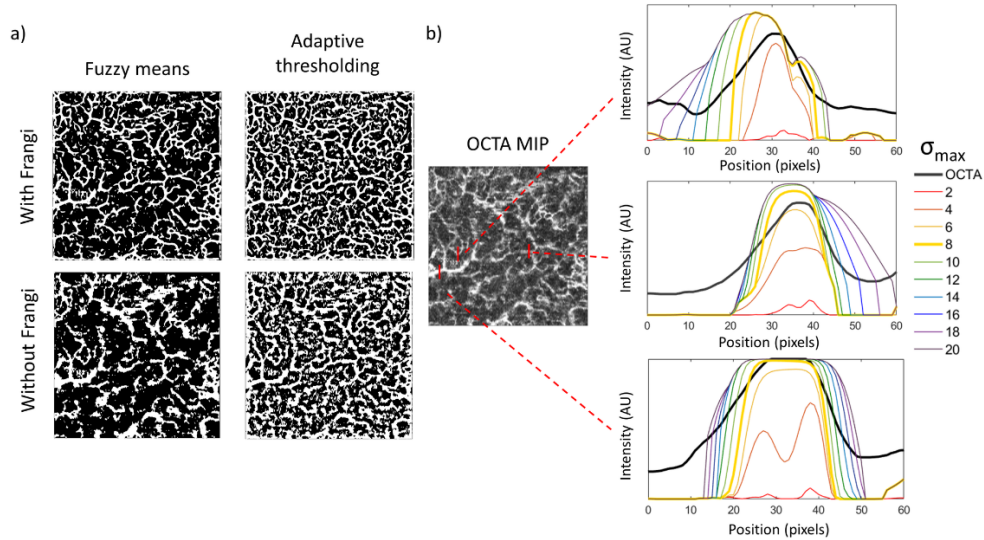

**Fig 4. Optimization of Frangi filter.** a) Indication of the impact of the Frangi filter on the interconnectivity of the vascular network after segmentation using two methods, as labeled; in both cases, $\sigma_{max}$ = 8. b) Demonstration of the impact of the Frangi filter on SNR and apparent vessel diameter for a range of $\sigma_{max}$ values. Red lines on the OCTA MIP image (center) mark three vessels with different diameters and OCTA and filtered line profiles in the graphs on the right as examples of the impact of the Frangi filter on apparent diameter in different scenarios. The OCTA MIP image dimensions are 5 mm x 5 mm, and the MIP comprises an axial range of 500 µm.

### 3.2 Validation of OCTA segmentation algorithms

A comparison of the performance of each of the optimized segmentation algorithms for three images with different vascular network densities and SNRs can be seen in the top part of **Fig 5**. In general, the built-in global thresholding algorithm within ImageJ does not accurately represent the complexity of the vascular network for *in vivo* OCTA MIP images of skin. It can be seen by inspection that the ImageJ global thresholding approach tends to underestimate the density of the vascular network and the resulting segmentation is poor. For images with high vessel density, such as the bottom row, all four of the remaining algorithms performed reasonably well; however, for images with lower VAD or poor SNR, ISODATA and adaptive thresholding tend to add additional structures within noisy but largely uniform regions of the image regardless of the kernel size specified. This means that these algorithms may have limited utility for images of skin but may still be useful for retinal images, which tend to have fewer uniform regions. The fuzzy thresholding and k-means algorithms replicate the vascular network in the binary mask over a larger variety of images. The five segmentation algorithms were also compared quantitatively using two metrics: VAD compared with manual calculation from the OCTA MIP image and network connectivity factor. The results of the quantitative comparison can be seen in the bottom of **Fig 5**. ISODATA and adaptive thresholding led to a high connectivity factor for all three images, which is not representative of the apparent differences in the images (higher CF does not necessarily correspond to better segmentation). K-means and fuzzy thresholding performed similarly on both VAD and connectivity factor and led to a VAD closer to the manual calculation, indicative of good segmentation. Based upon this analysis performed over a large range of images, we have determined that the fuzzy thresholding algorithm provides the most accurate segmentation for our OCTA MIP images of skin and is applicable to a wider variety of images than the other algorithms. As a result, we have only implemented the fuzzy thresholding and adaptive thresholding in the final version of OCTAVA. While adaptive thresholding did not perform well with our images, it is used commonly in retinal OCTA; we included it only as a basis for comparison for users who may be more familiar with the adaptive thresholding algorithm.

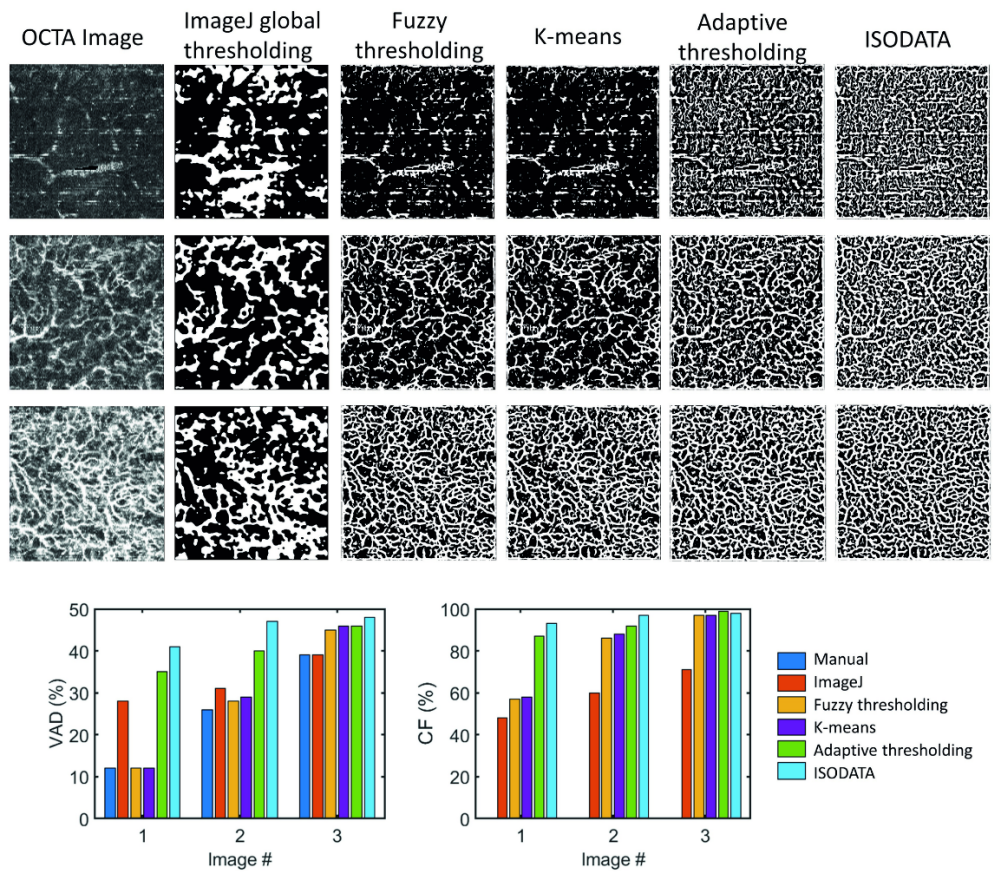

**Fig 5. Comparison of five segmentation algorithms for three images with different apparent VAD**. Top: binarized OCTA MIP images. Bottom: quantitative analysis. All images are OCTA MIP images comprising an axial range of 500 µm and with dimensions of 5 mm x 5 mm. VAD: vessel area density; CF: connectivity factor.

*3.3 Validation of vascular architecture metrics using microfluidic device and a simulated OCT MIP image*

To validate the accuracy of our generated metrics, we used two starting images: a conventional OCT *en face* image of a microfluidic device with known channel diameters; and a simulated OCT MIP image with hand-drawn structures mimicking a vascular network (**Fig 6**). The image of the microfluidic device was collected with a spectral domain OCT system described in [53]. The original OCT image depicted six small channels merging into one large channel. Although it conveys the advantage of known sizes, it was difficult to analyze this image since the pattern of the microfluidic device is not representative of a vascular network. Therefore, we synthesized a new image by copying and tiling the original image in a grid pattern to better mimic a network of vessels. For both images, the binary mask (**Fig 6b** and **g**) was generated using the fuzzy thresholding algorithm. Due to the bimodal distribution of vessels in the microfluidic image, the Frangi filter was not used; the performance of the Frangi filter is highly dependent on the range of vessel sizes within an image, so it is not possible to optimize the filter parameters for two distinct diameter ranges without distorting the image [59]. For the simulated OCTA MIP image, intensity variations were introduced along the simulated vessels so that the performance could be evaluated including the image enhancement step (without intensity variations, the Frangi filter would have no effect). The Frangi filter was used with a $\sigma_{max}$ value of 7 and was optimized using a procedure similar to that described in Section 3.1. The pixel size in each image (4 µm for the microfluidic image and 9.3 µm for the simulated OCTA MIP) determines the expected error bounds for diameter and length measurements.

The segmentation was evaluated by plotting the intensity of the OCT image and the binary mask along a vertical or horizontal line. The results demonstrate that the segmentation step does not impact the apparent channel diameters in most cases, although the Frangi filter may increase the diameter of small, low-intensity vessels depending on the chosen value of $\sigma_{max}$. This is further confirmed by the histogram of channel diameters for both images (**Fig 6e** and **j**). As expected, the histogram of diameters for the microfluidic device indicates a bimodal distribution with peaks at the correct values of 50 µm and 300 µm representing the small and large channels. Measured diameters are typically within the error bounds determined from the pixel size with a few exceptions. For example, some of the channels in the microfluidic device are measured to have larger diameters than expected due to the functioning of the algorithm at junctions between the small and large channels, as can be seen from the overlay image (**Fig 6c**). Someone manually characterizing the image might identify a node at the endpoint of each of the 50-µm channels before any broadening or change in direction indicating the junction with the larger channel; the automated characterization often includes part of this transition region as part of the smaller channels and places the node closer to the 300-μm channel. This can be seen most clearly in the top and bottom channels of each row. As a result, the measured average diameter of the 50-μm channel is larger than it would be if the transition region were identified as an independent segment. For comparison, the nodes identified by OCTAVA were taken to be the segment endpoints for the manual characterization of the network. The thickness maps **(Fig 6d** and **i)** demonstrate accurate diameter measurements throughout both images. The histograms of the lengths also match closely with the expected values (not shown).

The metrics generated by OCTAVA were compared with values calculated manually for both the microfluidic image and simulated OCTA MIP image (**Table 2**). The VAD was calculated manually using the mean intensity of the image as a global threshold. This is a valid approach for these images since there is a clear difference in intensity level between the network of channels and the background. Diameter and length measurements for the microfluidic images were measured manually in ImageJ in order to manually calculate VLD, mean and median diameter, and branchpoint density; for the simulated OCTA MIP image, exact diameter and length measurements were calculated based on *a priori* knowledge of the image structure. In all but one measurement, the relative difference between the automatic and manual metric calculations did not exceed 10%, although OCTAVA tends to overestimate the vessel diameter when compared with manual characterization based on the way it identifies nodes and vessel endpoints as described above. The automated method we use for measuring the vessel diameter (the local thickness algorithm in ImageJ) is commonly used for measuring the diameter of structures within an image (irrespective of the image content), however, it is difficult to replicate it exactly in manual analysis. This difficulty, in conjunction with the placement of nodes described above, and the limitation based on pixel size, accounts for the differences in the diameter measurements.

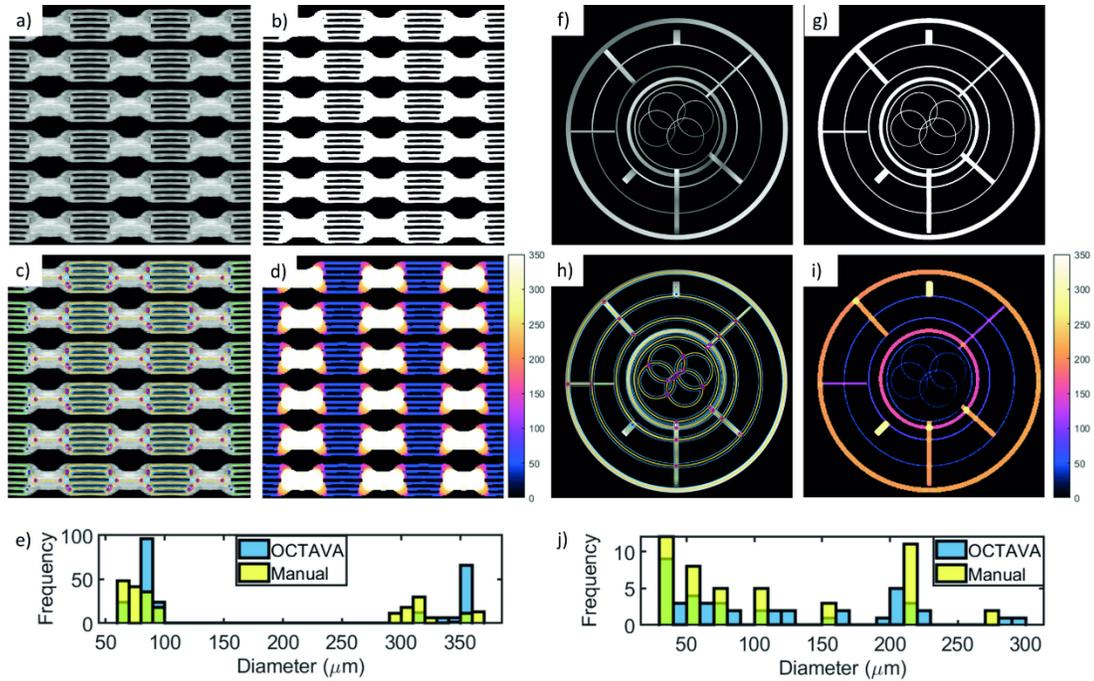

**Fig 6. Validation of image processing steps and metric results using images of a microfluidic device (a-e) and a simulated OCTA MIP image (f-j).** The composite image (a) is 4.32 mm x 4.32 mm with a pixel size of 4 μm. The binary mask (b) and overlay image (c) demonstrate the accuracy of the segmentation algorithm. The thickness map (d) demonstrates accurate measurement of the channel diameters throughout the image. The color bars in (d) and (i) indicate vessel diameter in μm. The green areas in the histogram indicate overlap between the blue and yellow bars. The same analysis was repeated for the simulated OCTA MIP image, which was assumed to be 10 mm x 10 mm with a pixel size of 9.3 μm. The overlay images (c, i) include a color-coded notation indicating different structures within the network: segments (yellow); branches (green); nodes (pink and blue circles); mesh regions (blue). (e) and (j) show the measured distribution of diameters in the microfluidic image and simulated OCTA MIP image, respectively. The larger channels in the microfluidic device are 300 μm and the smaller channels are 50 μm.

**Table 2. Comparison between OCTAVA and manual calculation of metrics.**

| Metric (unit) | Microfluidic device | | | Simulated OCTA MIP | | |
|---|---|---|---|---|---|---|
| | OCTAVA | Manual | Relative difference (%) | OCTAVA | Manual | Relative difference (%) |
| **Vessel area density (%)** | 54 | 54 | 0 | 13 | 14 | 7 |
| **Vessel length density (%)** | 2.16 | 2.16 | 0 | 1.1 | 1.1 | 0 |
| **Mean diameter (μm)** | 185 | 169 | 9 | 113 | 110 | 3 |
| **Median diameter (μm)** | 90 | 88 | 2 | 78 | 70 | 11 |
| **Branchpoint density (nodes/mm)** | 1.6 | 1.7 | 6 | 0.27 | 0.27 | 0 |
| **Fractal dimension** | 1.35±0.36 | | | 1.22±0.36 | | |
| **Mean tortuosity** | 0.07 | | | 0.15 | | |

### 3.4 Repeatability of OCTAVA metrics

Three repeated scans of the same skin location of eight study participants were used to validate the repeatability of OCTAVA metrics. The repeatability of each metric was calculated based on the within-subject standard deviation ($S_w$) method introduced by Bland and Altman [60]. The standard deviation of repeated measures for each subject was calculated, squared to obtain the variance for each subject, and then the square root of the average variance for all subjects gives the measurement error, $S_w$. The coefficient of repeatability (CR) is defined as $2.77 S_w$ and 95% confidence intervals (CI) as: CR$\pm 1.96 \cdot (S_w / \sqrt{2n(m-1)})$, where $n$ is the number of subjects and $m$ the number of measures for each subject [61]. **Table 3** and **Fig 7** summarize the outcome of the analysis. In **Table 3**, mean is the mean of all 24 measurements and measurement error (ME) is the mean of $S_w$ for all the participants. The factors that determine CR are related to differences in the raw OCTA images and the analysis. The repeatability of the software to obtain consistent results on the same image was also verified. If no changes are made to the parameters used to generate the binary mask, the algorithm will repeatably calculate identical metrics with no error on separate trials. All values are consistent with related analyses previously reported [58,62–64]. For example, Byers et al. have reported a mean vessel diameter of 53 μm and fractal dimension of 1.39 for the medial elbow of healthy volunteers [58].

**Table 3. Intrasession repeatability of vascular metrics obtained using OCTAVA.**

| Metric (unit) | Mean | ME[a] | CR[b] | 95% CI[c] |
|---|---|---|---|---|
| **Vessel area density (%)** | 38 | 2 | 5.5 | $4.8 - 6.2$ |
| **Vessel length density (%)** | 2.32 | 0.14 | 0.38 | $0.33 - 0.43$ |
| **Mean diameter (μm)** | 54.6 | 1.4 | 3.9 | $3.4 - 4.4$ |
| **Median diameter (μm)** | 56.2 | 1.8 | 5 | $4.4 - 5.7$ |
| **Branchpoint density (nodes/mm)** | 3.2 | 0.1 | 0.3 | $0.3 - 0.4$ |
| **Fractal dimension** | 1.38 | 0.01 | 0.02 | $0.02 - 0.02$ |

[a]ME: Measurement error; [b]CR: Coefficient of repeatability; [c]CI: Confidence intervals

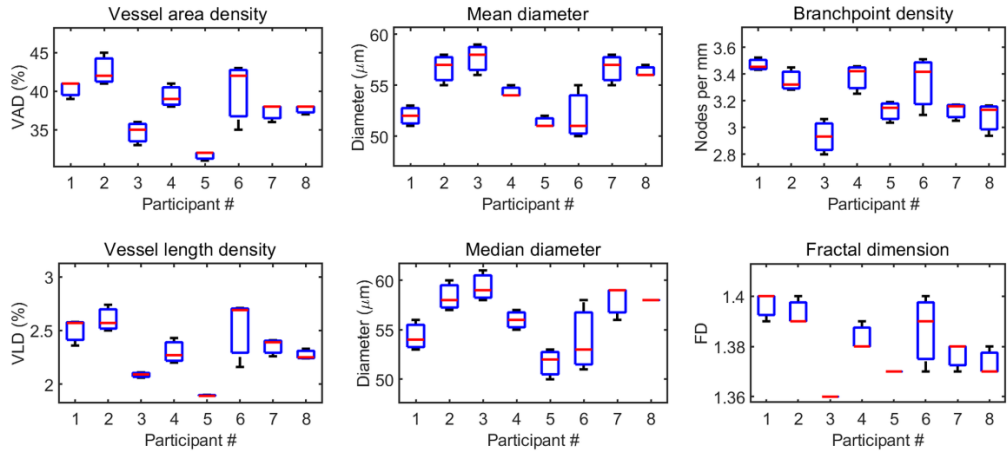

**Fig 7. Boxplots showing the repeatability of various vascular metrics obtained from selected skin datasets.** Black lines (limits on the boxes) represent the range of values obtained; red lines indicate the median value; blue boxes indicate the 25th and 75th percentiles. FD: fractal dimension.

*3.5 Application of OCTAVA to other OCTA instruments and imaging applications*

The performance of OCTAVA was further evaluated using several other published OCTA MIP images of skin and retina [54–56]. The results are presented in **Fig 8**. In all cases, the segmentation (**Fig 8b, f, and j**) and network identification (**Fig 8c, g, and k**) showed high-quality reproduction of vessel morphology. A heatmap was generated of the locally identified vessel diameters throughout the image and the range of measured diameters is within the expected range for each case. This outcome is notable given that the same software was used to process and analyze all three images. The selected images came from several published works demonstrating the applicability of the software on different instruments, imaging targets, and applications. Given the range of metrics implemented within OCTAVA, we were able to describe additional metrics of these images beyond those presented in the original publications.

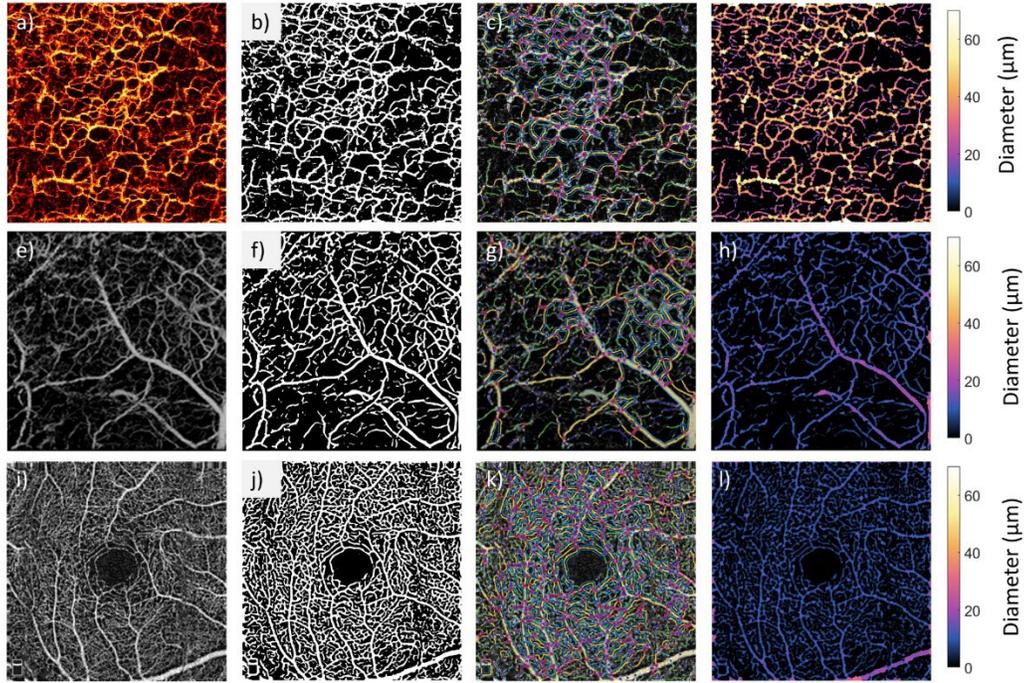

**Fig 8. OCTAVA applied to OCTA acquired with three different third-party instruments and imaging targets.** Top row: OCTA MIP image of human forearm acquired using a lab-built PS-OCT system. The image is 6.8 mm x 6.8 mm, and the MIP spans an axial range of 600 µm. Image reprinted with permission from[54]. Middle row: OCTA MIP image of mouse ear skin acquired with Telesto OCT system. The image is 2.5 mm x 2.5 mm and the axial span of the MIP is 1 mm. Reprinted with permission from [55]. Copyright (2021) American Chemical Society. Bottom row: OCTA MIP image of human retina acquired with RTVue XR Avanti. The image is 3 mm x 3 mm, and the axial span of the MIP is about 20 µm. Reprinted with permission from [56]. First column contains the original OCTA image; remaining columns show the evaluation steps using OCTAVA (left to right, the binary mask, overlay of skeleton, and map of vessel diameters). In all cases, the binary mask was generated using the Frangi filter ($\sigma_{max} = 8$ for top row and $\sigma_{max} = 6$ for middle and bottom rows) and fuzzy thresholding segmentation.

## 4. Discussion

While we do not propose any new algorithms here, we argue that standardization is the main attribute required to move the field forward. Many algorithms and processing workflows have already been proposed, usually proprietary to a research group or instrument vendor, to the detriment of open science. Overall, there is no generally accepted best workflow for segmenting and analyzing OCTA data [31,59], either overall, independent of the imaging target, or for a particular application. Our innovation comes from making informed choices through comparison of some of the best performing algorithms with the ambition of establishing a widely accepted and utilized standardized processing workflow suitable for all OCTA image types. To maximize its ease of uptake, we have packaged our proposed solution as an open-source, easy-to-use toolbox which is accessible to users at all levels of technical competency. This is important because it can provide a starting point for new entrants to the field of OCTA and is accessible to clinicians and end users while still being customizable for users seeking to adapt the software to their own needs. All these features together represent the power of OCTAVA.

## 4.1 Workflow optimization and comparison with other software packages

The vessel enhancement and segmentation steps are key for accurate identification of vessels, and different segmentation algorithms produce significant variation in measured metrics. With this in mind, we will briefly compare OCTAVA with other reported software used for analysis of OCTA images. This comparison will contextualize OCTAVA as a step towards a standardized workflow that could be used for all OCTA images.

### Vessel enhancement

In angiography more generally, vessel enhancement (filtering) is often used to improve the visibility of vessels. Here, we briefly justify our choice of the Frangi filter for vessel enhancement. Hessian-based filtering (including the Frangi filter) has been commonly used in magnetic resonance angiography, ultrasound angiography, and photoacoustic angiography [65–67]. Specifically for OCTA, there are many possible approaches, each with advantages and disadvantages, including Hessian-based filtering, rod-filtering, top-hat filtering, optimally oriented flux enhancement, Gabor filtering, weighted symmetry filtering, and active shape models [25,59]. Among these approaches, Hessian-based filtering has been shown to result in the best quality of final image [25]. Other works have employed a Frangi filter to improve visibility of vessels in OCTA images of the retina [48], choroid [68], and skin [40]. Despite its widespread use, the Frangi filter has also been criticized in some studies for introducing errors in the vessel architecture, either by missing vessels rendered in the image with low SNR or generating spurious vessels depending on the structure of the background noise [69,70]. Even using the multiscale Frangi filter approach, it has been shown that the results are highly dependent on the range of vessel sizes within the image [71]. Our study of Frangi filter behavior with OCTA images of the skin demonstrates that for $\sigma_{max}$ values in the range of 1-8, vessel dilation is minimal. However, parameter optimization must be adapted for each imaging application, which is a limitation on standardization. We finally note that modifications to this algorithm to further optimize it for OCTA images represents a possible further improvement [71].

### Filling gaps in vessels: Manual curation vs automation

Improving network connectivity by extrapolation is another issue that must be considered as a potential image pre-processing step. It is expected in a real microvascular architecture, that all vessels are fully integrated in a network. Therefore, breaks in vessels and disconnected endpoints far from the image boundaries do not accurately represent vessel morphology. AngioTool (an open source tool optimized for confocal fluorescence images), among other reported tools, has incorporated the ability to extrapolate to fill gaps in vessels to improve connectivity [21]. A required trade-off in such approaches is the acceptable degree of image manipulation balanced against reasonable expectations about the vessel network. For example, the discontinuity may be due to a change in direction or depth of the vessel, thus, adding the connection may be erroneous. Furthermore, since OCTA can only visualize perfused vessels with flow rate above a minimum threshold, non-patent microvessels and obstructions appear as discontinuities and could be overlooked if the gaps are filled automatically [72]. At this stage, we have chosen not to incorporate automated gap-filling to avoid ambiguous manipulation of images. In some cases, where added connection is justified, the user may use the manual curation option to add connections. Although, to date we have found that (depending on the vessel density) these cases cause minimal change in the values of the overall metrics (see **Section 4.2**).

### Image segmentation: computer-learning vs traditional algorithms

Some recent studies have investigated the use of deep-learning algorithms to improve vessel segmentation [25,27]. For example, Stefan and Lee have compared their deep learning segmentation algorithm (optimized for images of the mouse brain) with global thresholding and adaptive thresholding for volumetric OCTA image processing [25]. Although they demonstrated good performance, deep-learning and other computer-learning-based approaches for image segmentation are incompatible with the development of a standardized workflow since computer learning algorithms must be trained using a particular set of images.

Several recent studies in retinal OCTA have evaluated the performance of various traditional segmentation algorithms and found significant variability between them [17,28–30]. Of these, the fuzzy means algorithm has shown promise for imaging in the choroid and retina [68,73]. Combined with the results presented here in skin, the fuzzy means algorithm is beginning to stand out as a good option for standardizing the processing workflow.

Choice and definition of metrics

Many metrics have been used for characterizing microvasculature; an overview of common metrics can be found in [43]. One of the challenges impeding widespread adoption of clinical OCTA is the wide variability in how these metrics are defined and calculated. This is compounded by the fact that most software available for analyzing vascular networks does not allow access to the raw data used to generate those metrics. For example, AngioTool is commonly adapted for OCTA images but only provides a summary of metrics and does not provide information about individually identified vessels [21]. By contrast, OCTAVA makes the measurements from all identified vessel segments available to enable further statistical analysis and generation of additional metrics. Similarly to [25], OCTAVA uses a graphical approach to present the outcome of the quantitative analysis through histograms of diameter, length and tortuosity, which allows for further analysis of the mechanisms driving changes in metrics which characterize the image as a whole, such as vessel density and mean diameter.

## 4.2 OCTA image artifacts and the impact on segmentation and metrics

One of the main limiting factors in the application of OCTAVA is the presence of motion artifacts. Motion artifacts are a main limiting factor in any segmentation-based analysis because the algorithm does not differentiate artifacts from vessels. Casper et al. have recently proposed a segmentation technique which incorporates a refinement step to help correct for defocus and blurring caused by motion, which may prove helpful [70]. In any case, if motion artifact exceeds a certain level (such as the example in **Fig 9a and b**), there is no choice but to exclude the images from analysis. Thus, due attention should be paid to stabilize the patient/instrument interface to minimize motion artifact in the first place. For images with infrequent motion artifacts, OCTAVA allows for improperly identified features to be manually excluded from the analysis using manual curation. One example of this and the impact it has on the image metrics is shown in **Fig 9c-h.** The impact of the motion artifact correction on the metrics depends on the overall information content of the image. However, we found that in cases where it was feasible to manually remove motion artifacts, the impact on the value of the metrics was minimal. Manual curation can also be used to remove artifacts from other sources, such as floaters in the eye or edema.

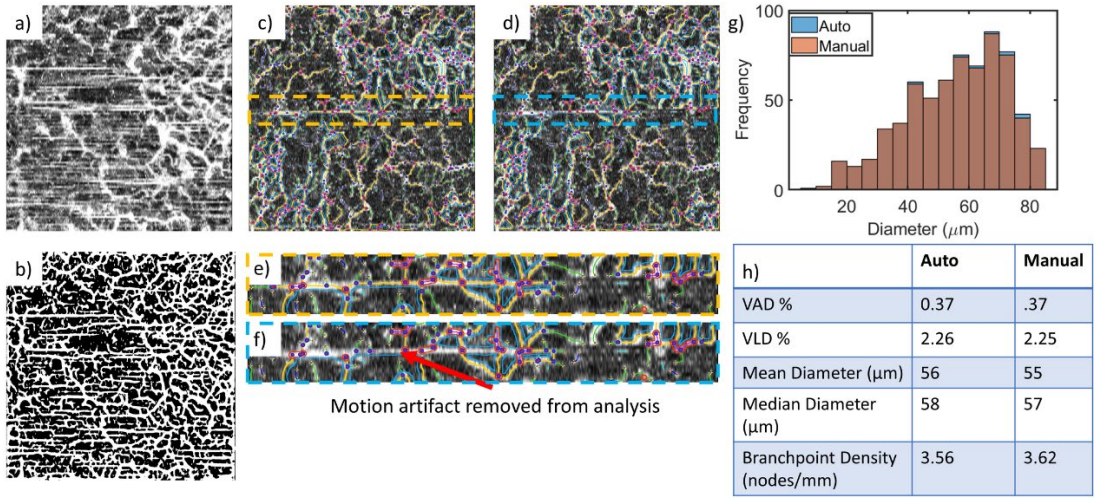

**Fig. 9. Motion artifacts, poor-quality images, and manual curation.** Images with significant motion artifact (a) must be removed from the analysis since the segmentation algorithm cannot distinguish between the vessels and motion artifacts; the resultant segmented OCTA MIP (b) does not accurately represent the microvascular architecture. For images with minimal motion artifacts, such as the images in (c) and (d), the individual motion artifacts can be removed manually. (e) and (f) show insets of (c) and (d), respectively, demonstrating the removal of the motion artifact using manual curation. Depending on the VAD, individual motion artifacts likely have little impact on the generated metrics, as shown in the histogram of diameters (g) and the table of generated metrics with and without manual curation (h). Note that in (g), the columns representing the results with manual curation are always equal in height or shorter than the automatically generated values since no vessels were added by manual curation in this instance. Square OCTA MIP images in (a-d) are 5 mm x 5 mm and comprise an axial range of 500 µm. The images in (e) and (f) are 0.65 mm x 5 mm.

## *4.3 Future directions*

We are considering several improvements to the segmentation and analysis workflow in future versions of OCTAVA. One consideration is the ability to further segment images based on vessel diameter, which some studies suggest could bring extra diagnostic power [27,74,75]. Further segmentation based on vessel diameter may be used to differentiate between different vessel groups (*e.g.,* venules, arterioles, and capillaries). It is known that different eye diseases and severity stages can affect the arterioles and venules differently [76,77], so it would be useful to be able to distinguish between them.

Another consideration is 3D-based analysis of the OCTA data. One inherent limitation of analyzing MIPS is that it is impossible to discern between two vessels that intersect or cross at different depths [43]. As shown by Sarabi et al., 3D analysis enables better visualization and captures details of vascular geometry not seen in 2D analysis [78]. However, challenges persist in 3D analysis of OCTA images due to projection artifacts inherent in OCTA imaging [79]. Taking into consideration how common and clinically applicable 2D analysis is, we think a standardized image processing workflow in 2D is still a significant step towards facilitating open data sharing and cross-comparison and pooling of data from different studies.

While our initial optimization has focused on skin, we recognize the wide application of OCTA imaging to retinal analysis, which requires the ability to measure several additional metrics including the area and perimeter of the foveal avascular zone, and the vessel density within several subregions of the image [80]. Retinal analysis will be delivered in the next version of the software.

Finally, one of the goals of our research is to create multi-source image databases available to be mined to find better biomarkers of health and disease. For example, with the creation of large databases, we will provide enhanced opportunity for rapidly emerging AI-based image classification schemes to identify biomarkers.

## 5. Conclusion

We have developed a fully integrated and easy-to-use software tool, OCTAVA, for quantitative analysis of OCTA MIP images based on a range of metrics. We have demonstrated that quantification of the metrics with OCTAVA is accurate and repeatable. We have demonstrated OCTAVA's versatility by showing it can be applied to OCTA images acquired from a range of instruments on different imaging targets. We believe OCTAVA is an important step towards standardization of the processing workflow for OCTA data. Such standardization will lead to wide-ranging benefits in its application.

**Funding.**

GU received funding from the IPRS (University of Western Australia) and the Rank Prize Covid-19 response fund. RM and AD received funding from the Doctoral College (University of Surrey). CH and DMS received competitive internal funding from the University of Surrey.

**Acknowledgments.** DDS and DMS acknowledge the efforts of many colleagues working on OCTA, particularly K. Karnowski, Q. Li and P. Gong.

**Disclosures.** None

**Data availability.** Data underlying the results presented in this paper are not publicly available at this time but may be obtained from the authors upon reasonable request. Source code is available at https://github.com/GUntracht/OCTAVA. A compiled MATLAB app or standalone version of the software is available upon request.

2012;18: 061213. doi:10.1117/1.jbo.18.6.061213